\documentclass[aps,prc,superscriptaddress,twocolumn,nofootinbib]{revtex4}

\usepackage{amsmath,amssymb}
\usepackage{mathtools}
\usepackage{bbold}
\usepackage{braket}
\usepackage{graphicx,bm}
\usepackage{float}
\usepackage{multirow}
\usepackage{color}

\allowdisplaybreaks

\newcommand{\msun}{M_{\odot}}
\newcommand{\pt}{\partial}

\begin{document}

\title{Symmetry energy and neutron star properties\\
constrained by chiral effective field theory calculations}

\author{Yeunhwan Lim}
\email[E-mail:~]{ylim@yonsei.ac.kr}
\affiliation{Department of Physics, Yonsei University, Seoul 03722, South Korea}

\author{Achim Schwenk}
\email[E-mail:~]{schwenk@physik.tu-darmstadt.de}
\affiliation{Technische Universit\"at Darmstadt, Department of Physics, 64289 Darmstadt, Germany}
\affiliation{ExtreMe Matter Institute EMMI, GSI Helmholtzzentrum f\"ur Schwerionenforschung GmbH, 64291 Darmstadt, Germany}
\affiliation{Max-Planck-Institut f\"ur Kernphysik, Saupfercheckweg 1, 69117 Heidelberg, Germany}

\begin{abstract}
We investigate the nuclear symmetry energy and neutron star properties using a Bayesian 
analysis based on constraints from different chiral effective field theory calculations using
new energy density functionals that allow for large variations at high densities. Constraints
at high densities are included from observations of GW170817 and from NICER. In particular, we
show that both NICER analyses lead to very similar posterior results for the symmetry
energy and neutron star properties when folded into our equation-of-state framework.
Using the posteriors, we provide results for the symmetry energy and the slope
parameter, as well as for the proton fraction, the speed of sound, and the central density
in neutron stars. Moreover, we explore correlations of neutron star radii with the pressure
and the speed of sound in neutron stars. Our 95\% credibility ranges for the symmetry energy 
$S_v$, the slope parameter $L$, and the radius of a 1.4$\msun$ neutron star, $R_{1.4}$, are 
$S_v=(30.6\text{--}33.9)$\,MeV, $L=(43.7\text{--}70.0)$\,MeV, and $R_{1.4}=(11.6\text{--}13.2)$\,km. Our analysis
for the proton fraction shows that larger and/or heavier neutron stars are more likely to
cool rapidly via the direct Urca process. Within our equation-of-state framework a maximum
mass of neutron stars $M_{\rm max}>2.1\msun$ indicates that the speed of sound 
needs to exceed the conformal limit.
\end{abstract}
	
\maketitle

\section{Introduction}

Understanding dense matter is a central challenge in nuclear physics and astrophysics. In nature, dense matter exists in the cores of neutron stars under extreme neutron-rich conditions. The properties of neutron-rich matter around nuclear densities are described by the nuclear symmetry energy and its density dependence. While there have been impressive constraints from nuclear theory, nuclear experiments, and astrophysics (see, e.g., Refs.~\cite{LattimerLim,Dris21ARNPS,Huth21,Essick21PRC}), more precise determinations of the symmetry energy and its slope parameter $L$ at saturation density, $n_0 = 0.16$\,fm$^{-3}$, are still an open problem. 

From the theoretical side, the symmetry energy is best constrained by controlled calculations of the equation of state (EOS) of neutron matter based on chiral effective field theory (EFT) interactions~\cite{Hebe10nmatt,Tews13N3LO,Carb13nm,Hage14ccnm,Lynn16QMC3N,Holt17,Dris19MCshort,Jiang20,Kell23ANM}. This yields values for the symmetry energy $S_v$ at saturation density and the $L$ parameter in the ranges of $S_v = (30\text{--}35)$\,MeV and $L = (35\text{--}70)$\,MeV. However, to describe the EOS to all densities in neutron stars requires extensions beyond the reach of chiral EFT calculations. To this end, different extensions, such as piecewise polytropes~\cite{Hebe13ApJ}, speed-of-sound based parametrizations~\cite{Tews18cs,Greif19cs}, nonparametric Gaussian processes~\cite{Landry19GP}, or nuclear energy-density functionals (EDFs) have been used (see, e.g., Ref.~\cite{Lim18}).

Recently, new EDFs for the nuclear EOS were introduced by Huth {\it et al.}~\cite{Huth21}, and have the advantage of providing high-density extrapolations that are consistent with causality and with a maximum of the speed of sound. These functionals allow for EOS calculations for the broad ranges of conditions reached in core-collapse supernovae and neutron star mergers. In this work, we use these new EDF EOSs to constrain the symmetry energy and neutron star properties based on a prior informed by chiral EFT calculations of neutron matter.

From the astrophysics side, the strongest constraint on the nuclear EOS comes from the observation of heavy two-solar-mass neutron stars~\cite{Demo10ns,Anto13ns,Fonseca21}. Moreover, the heaviest well measured neutron star, PSR J0740+6620, was recently also observed by NICER to provide constraints on its radius~\cite{Riley21,Miller21}. In addition, NICER observed the mass and radius of a typical-mass neutron star, PSR J0030+0451~\cite{Riley19,Miller19}. The NICER analyses for both neutron stars by Riley {\it et al.}~\cite{Riley19,Riley21} and by Miller {\it et al.}~\cite{Miller19,Miller21} give different mass-radius posteriors, but agree within their uncertainties. The differences in the posteriors are reduced by including realistic assumptions for the EOS, and in this work we explicitly show that in our EDF EOS ensembles the results from both NICER analyses are very similar. In addition to the NICER constraints, we include in our Bayesian inference the tidal deformability information from GW170817 inferred by results from LIGO/Virgo~\cite{LIGO19PRX}. Using the chiral EFT informed priors with the astro posteriors, we provide results for the symmetry energy and neutron star properties.

This paper is organized as follows. In Sec.~\ref{sec:edf} we introduce our EOS framework using the new EDFs from Huth {\it et al.}~\cite{Huth21}. These are fit to a range of chiral EFT calculations of neutron matter. Building on this EOS prior, we include constraints at high densities from observations of GW170817 and from NICER using a Bayesian analysis. In Sec.~\ref{sec:neutronstar}, we investigate the posterior distributions for the symmetry energy and the slope parameter, as well as for the proton fraction, the speed of sound, and the central density in neutron stars. Moreover, we explore correlations of neutron star radii with the pressure and the speed of sound in neutron stars. Finally, we summarize our results and conclude in Sec.~\ref{sec:summary}.

\section{Equation-of-state framework}
\label{sec:edf}

The EOS describes the energy density and pressure of matter for given baryon density, composition, and temperature. Since we focus on cold neutron stars, we consider zero temperature. For a given EOS, the mass and radius of neutron stars follow by solving the Tolman-Oppenheimer-Volkoff (TOV) equations~\cite{Tolm39TOV,Oppe39TOV}. Our starting point will be the EOS of homogeneous matter, which we constrain by empirical ranges of the properties of symmetric nuclear matter around saturation density and by neutron matter calculations. Based on each EOS, we calculate consistently the structure of the neutron star crust.

Since neutron stars are extremely neutron rich with proton fractions $\approx$$5\%$, the most important constraints for the EOS come from neutron matter calculations. In this work, we focus on neutron matter calculations based on chiral EFT interactions, which has the advantage that chiral EFT predicts consistent many-body interactions and enables systematic uncertainty estimates based on the EFT expansion~\cite{Hebe15ARNPS,Dris21ARNPS}. Neutron matter has been calculated based on chiral two- and three-nucleon interactions using many-body perturbation theory (MBPT)~\cite{Hebe10nmatt,Tews13N3LO,Holt17,Dris19MCshort,Kell23ANM}, quantum Monte Carlo (QMC) methods~\cite{Geze13QMCchi,Lynn16QMC3N}, self consistent Green's function (SCGF) methods~\cite{Carb13nm}, and coupled cluster (CC) theory~\cite{Hage14ccnm,Jiang20}. These calculations are able to include all interactions up to next-to-next-to-next-to-leading order (N$^3$LO)~\cite{Tews13N3LO,Dris19MCshort,Kell23ANM} and include uncertainty estimates from the EFT truncation~\cite{Dris19MCshort,Dris20PRL,Kell23ANM,Jiang20}.

\subsection{Energy density functionals}

To extend the EOS to high density we use nonrelativistic EDFs, which depend on the baryon number density $n$
and proton fraction $x$ of uniform matter. The baryonic energy density $\varepsilon(n,x)$ is expressed as
\begin{align} 
\varepsilon(n,x) &= \frac{1}{2m_N} \, \tau_n(n,x) + \frac{1}{2m_N} \, \tau_p(n,x) \nonumber \\[1mm]
& \quad+ (1-2x)^2 f_n(n) + [1-(1-2x)^2]f_s(n) \,,
\label{eq:EDF1}
\end{align}
where $\tau_n/2m_N$ and $\tau_p/2m_N$ are the neutron and proton kinetic densities, with nucleon mass $m_N$. It was shown that the dependence on isospin asymmetry is to a very good approximation quadratic~\cite{Dris16asym,Somasundaram21}, with the dominant nonquadratic contributions stemming from the kinetic densities, so that Eq.~\eqref{eq:EDF1} provides a very good approximation for asymmetric nuclear matter. The functionals $f_n(n)$ and $f_s(n)$ can be chosen to satisfy the constraints from neutron matter calculations 
and symmetric nuclear matter properties, respectively.

For the interaction density functionals, we take the form introduced recently by Huth {\it et al.}~\cite{Huth21},
\begin{equation}
f_n(n) = \sum_{j=0}^3 \frac{a_j n^{2+j/3}}{d_j + n^{(j+1)/3}} \,, \quad
f_s(n) = \sum_{j=0}^3 \frac{b_j n^{2+j/3}}{d_j + n^{(j+1)/3}} \,,
\label{eq:EDF2}
\end{equation}
where $a_j, b_j$ are fit parameters and $d_j =  d \, \mathrm{fm}^{-1-j}$ with parameter $d=3$~\cite{Huth21}. This corresponds to an expansion of the interaction energy density in powers of the Fermi momentum $k_{\rm F} \sim n^{1/3}$, and the denominator ensures that the interaction part becomes proportional to $n^{5/3}$ at higher densities. Note that without the denominator, the interaction part generally causes the speed of sound to exceed the speed of light beyond some baryon density. For a detailed discussion of these new functionals and the parameter choices, see Ref.~\cite{Huth21}.

\begin{figure}[t]
	\centering
	\includegraphics[width=0.9\columnwidth]{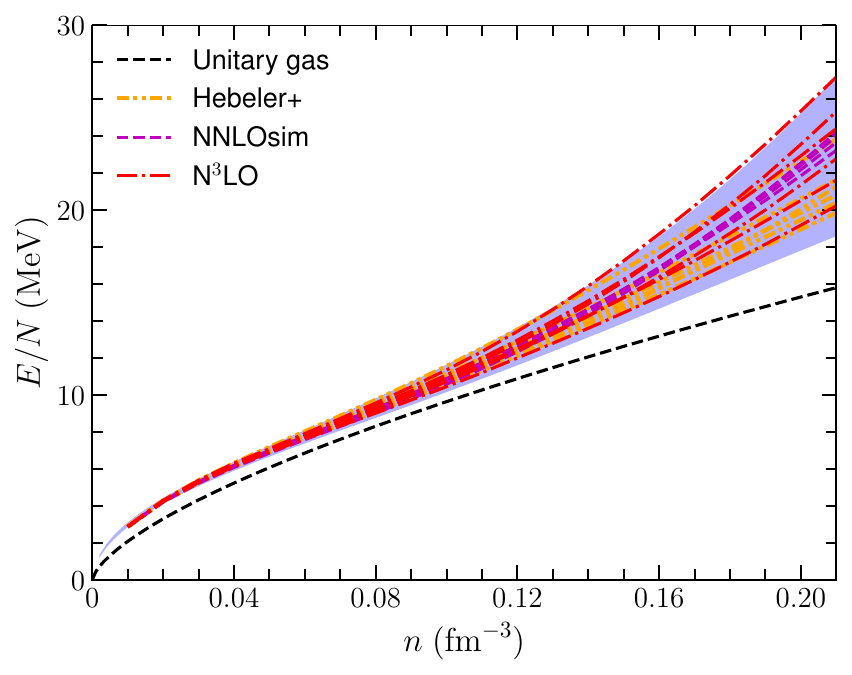}
	\caption{Neutron matter energy per particle $E/N$ as a function of number density $n$ based on the MBPT calculations from Ref.~\cite{Dris19MCshort} using different chiral $NN+3N$ Hamiltonians (labeled Hebeler+, NNLOsim, and N$^3$LO; for details see text). The band shows the $95\%$ credibility region modeling the different MBPT results with the EDF ensemble used in this work (based on the $k_{\rm F}$ expansion and $d=3$). For comparison we also show the unitary gas constraint~\cite{Tews17}.}
	\label{fig:eftpnmstat}
\end{figure}

\begin{figure}[t]
	\centering
	\includegraphics[width=0.9\columnwidth]{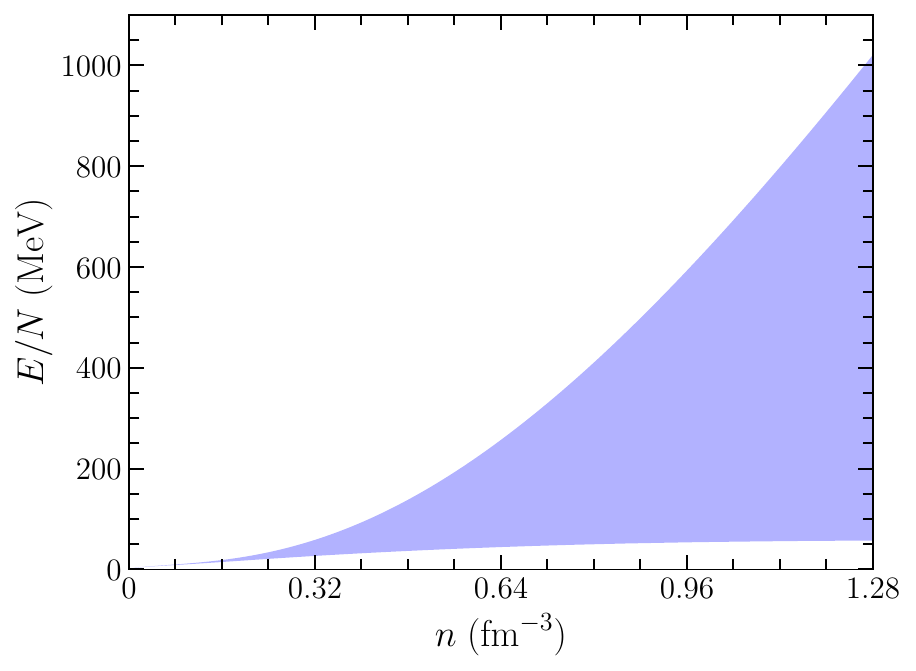}
	\caption{Same as Fig.~\ref{fig:eftpnmstat} but for the $95\%$ credibility region of the EDF ensemble used in this work (based on the $k_{\rm F}$ expansion and $d=3$), but extended up to $8n_0$.}
	\label{fig:eftpnmstat_high}
\end{figure}

\subsection{Constraints from neutron matter calculations based on chiral effective field theory}

For neutron matter constraints we use the MBPT calculations from Ref.~\cite{Dris19MCshort} based on different chiral $NN+3N$ Hamiltonians, including the Hebeler+ interactions~\cite{Hebe11fits}, the NNLOsim potentials~\cite{Carl15sim}, as well as the N$^3$LO 450\,MeV and 500\,MeV uncertainty bands~\cite{Dris19MCshort} (using the $NN$ Entem-Machleidt-Nosyk (EMN) interactions~\cite{Ente17EMn4lo}). The different neutron matter results and their uncertainties are given by the individual lines shown in Fig.~\ref{fig:eftpnmstat}. We use the individual lines to fit the $a_j$ of the EDF for neutron matter, $f_n(n)$ in Eq.~(\ref{eq:EDF2}), based on the $k_{\rm F}$ expansion and $d=3$.

The $b_j$ of the corresponding symmetric matter part, $f_s(n)$, are determined from empirical properties. We fit to the binding energy $E/A(n_0)=-16$\,MeV at saturation density $n_0 = 0.16\,\text{fm}^{-3}$, the incompressibility $K=235$\,MeV, with $K = 9 n^2 \pt^2(E/A)/\pt n^2(n_0,x=1/2)$, and the skewness $Q = -300$\,MeV, with $Q = 27 n^3 \pt^3(E/A)/\pt n^3(n_0,x=1/2)$. These values are extracted from Skyrme EDFs and constraints for nuclear matter properties~\cite{Dutra12PRC}; see also Ref.~\cite{Lim19}.  Since neutron star properties are not very sensitive to symmetric nuclear matter, we do not vary all nuclear matter properties, but only explore the most uncertain value of $Q$ in the following, see Sec.~\ref{protonfraction}.

The uncertainties in our EDF EOSs are reflected in the covariance matrix of $\vec{x}=(\vec{a}, \vec{b})$, defined as
\begin{equation}\label{eq:cov}
C_{jk} = \frac{1}{\sum_i w_i} \sum_i w_i (x_j^i -\langle x_j \rangle)(x_k^i -\langle x_k \rangle) \,,
\end{equation}
where  $x_j^i$ is the set of fit parameters $(a_j, b_j)$ for the $i$th individual EOS, $\langle x_j \rangle$ represents the average of $x_j$, and $w_i$ is the weight for each EOS. Since we do not vary the symmetric nuclear matter properties, in this work $C_{jk}$ is a $4 \times 4$ matrix for the $a_j$ from the neutron matter EOSs only. In the initial set given by the 17 neutron matter EOSs, the weights are $w_i=1$, but when we implement Bayes statistics and inferences, $w_i < 1$. With the average $\langle x_j \rangle$ and the covariance matrix $C_{jk}$, a multivariate normal distribution can be used to generate an EOS ensemble based on our EDF EOSs. We note that the statistical uncertainties from this EOS ensemble have of course a prior sensitivity to the initial set of individual EOSs.

The resulting EDF EOS ensemble based on the multivariate normal distribution is shown in Fig.~\ref{fig:eftpnmstat} with the $95\%$ credibility region in comparison to the individual EOSs based on MBPT calculations of neutron matter. The ensemble is based on 100,000 EOSs generated using the EDF, Eqs.~(\ref{eq:EDF1}) and~(\ref{eq:EDF2}), from the average $\langle x_j \rangle$ and the covariance matrix $C_{jk}$ based on the individual neutron matter MBPT EOSs. The agreement between the band and the individual lines in Fig.~\ref{fig:eftpnmstat} indicates that the EDF EOS ensemble employed in this work can generalize chiral EFT results within their uncertainties. Moreover, we compare the EDF EOS ensemble to the unitary gas constraint~\cite{Tews17} and observe in Fig.~\ref{fig:eftpnmstat} that this is nicely fulfilled by our EOSs. 
For completeness, we show in Fig.~\ref{fig:eftpnmstat_high} the prior of our EDF ensemble extended  up to $8n_0$.

\subsection{Bayesian modelling}

We incorporate the astrophysics constraints on the EOS by applying Bayes' theorem, from which the posterior distribution results from the combination of the prior and likelihood,
\begin{equation}
P(\vec{a} \vert D) = \frac{P(D|\vec{a})P(\vec{a})}{\int d\vec{a} \, P(D|\vec{a})P(\vec{a})}\,.
\end{equation}
Here, $P(\vec{a})$ represents the EOS prior given by the EDF parameter space obtained from the neutron matter calculations and symmetric nuclear matter properties, and $D$ stands for the astrophysical data so that $P(D|\vec{a})$ is the likelihood or conditional probability 
of obtaining $D$ for a given EDF with parameter set $\vec{a}$. 
In our study, we include the astrophysical observations of GW170817 and results from NICER
to constrain the EOS at higher densities.

For the NICER mass-radius constraints for PSR J0030+0451 and PSR J0740+6620 we consider separately either the Amsterdam analysis of Riley {\it et al.}~\cite{Riley19,Riley21} or the Illinois/Maryland analysis of Miller {\it et al.}~\cite{Miller19,Miller21}. The heaviest neutron star mass of $2.08 \pm 0.07,\msun$~\cite{Fonseca21} is thus directly implemented through the NICER $M\text{--}R$ information of PSR J0740+6620. Folding in the NICER constraints based on our prior leads to the likelihood for the EDF parameters~\cite{Lim2020n,Lim2022f}
\begin{align}
P(\text{NICER}|\vec{a}) &= \int  dM \, P(M ) 
 P(\text{NICER} \vert M, R(M,\vec{a})) \,,
\end{align}
where $P(M)$  is the probability domain of the NICER analysis and $P(\text{NICER} \vert M, R(M,\vec{a}))$ is the likelihood of each of the two NICER results for a given EDF EOS with parameter set  $\vec{a}$~\cite{Landry2020PRD}.
The integral is carried out be discretizing the $M-$ space, summing over all bins which are passed by the $M(\vec{a})\text{--}R(\vec{a})$ relation, and weighting those bins with the NICER posterior for each of the sources successively.

In addition to NICER, we use the tidal deformability information from GW170817 inferred by LIGO/Virgo~\cite{LIGO19PRX},
\begin{multline}
P(\text{LIGO}|\vec{a}) = \int dM_1 dM_2 \, P(M_1,M_2) \nonumber \\ 
\times P(\text{LIGO} \vert M_1 M_2 \Lambda_1(M_1,\vec{a}) \Lambda_2(M_2,\vec{a})) \,,
\end{multline}
where $P(M_1,M_2)$ is the probability domain from the LIGO/Virgo analysis and
$P(\text{LIGO} \vert M_1 M_2 \Lambda_1\Lambda_2)$ is the likelihood of the LIGO/Virgo analysis a given EDF EOS with parameter set  $\vec{a}$~\cite{Landry2020PRD}.
We assume that the NICER and GW170817 analyses are independent of each other so that, combining both constraints, the likelihood is given by
\begin{equation}
P(D\vert \vec{a}) = P(\text{NICER}|\vec{a}) P(\text{LIGO}|\vec{a}) \,.
\end{equation}
Multiplying the combined likelihood with the prior $P(\vec{a})$ and a normalization constant considering the integral in the denominator, we obtain the posterior distribution $P(\vec{a} \vert D)$ for a given EDF EOS with parameter set $\vec{a}$.

\section{Results}
\label{sec:neutronstar}

Next we present our results for the properties of neutron stars and the symmetry energy based on the EOS framework developed in the previous section. This combines the information from neutron matter based on chiral EFT interactions, with empirical properties of symmetric nuclear matter, as well as astrophysical constraints from GW170817 and NICER using a family of EDFs for nucleonic matter. Since matter in neutron stars is very neutron-rich, we have focused more on the propagation of the theoretical uncertainties in our knowledge of neutron matter. An advantage of our EOS framework is that we use the same EDF to construct the crust and core EOSs for neutron stars. In the following, we present our results for the neutron star mass and radius, the proton fraction, the speed of sound, and the central density in neutron stars. We also provide results for the symmetry energy and the slope parameter and explore correlations of neutron star radii with the pressure and the speed of sound in neutron stars. 

\subsection{Mass-radius relation}

\begin{figure}[t!]
	\centering
	\includegraphics[width=0.85\columnwidth]{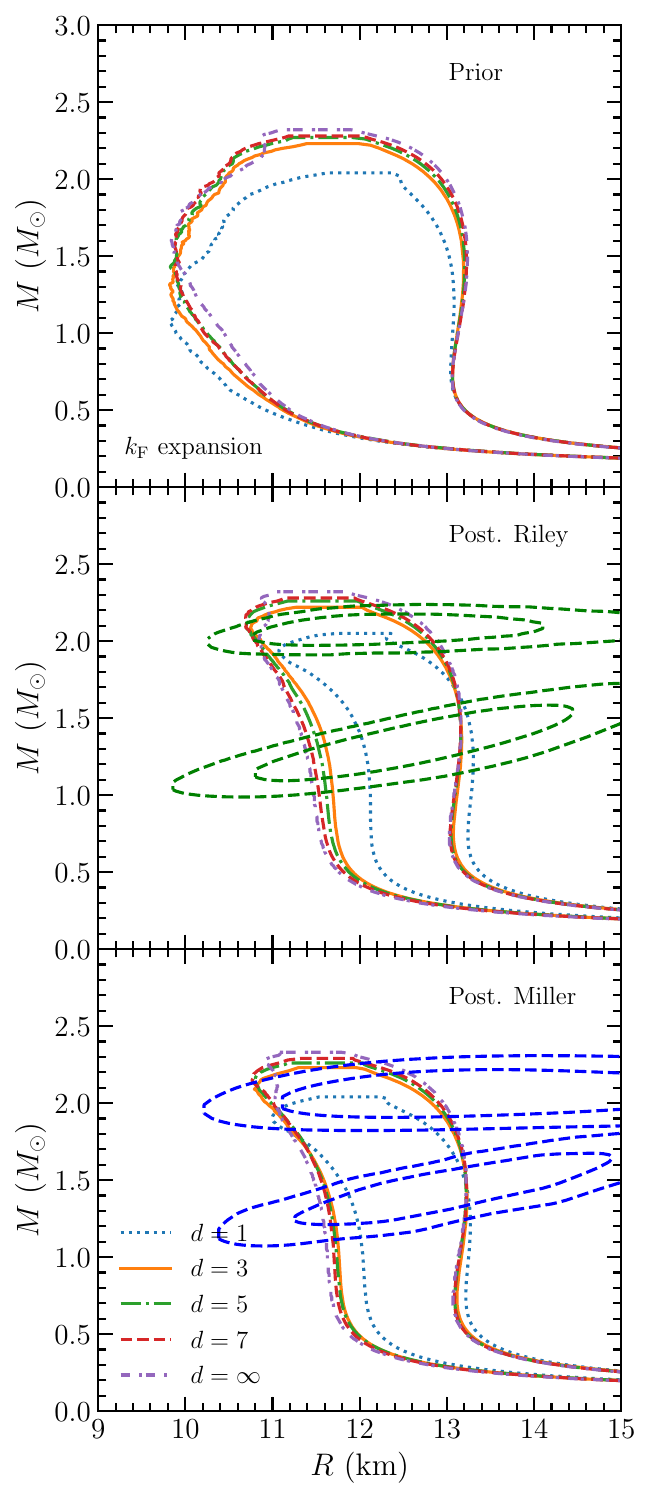}
	\caption{Top panel: 95\% credibility regions for the neutron star mass $M$ and radius $R$ generated from the multivariate normal distribution for the EDF EOSs with the $k_{\rm F}$ expansion using different $d$ values. Middle panel: Posterior distribution based on the top panel prior and including astrophysics information from GW170817 and NICER (dashed 95\% and 68\% contours from the Amsterdam analysis, Riley {\it et al.}~\cite{Riley19,Riley21}). Bottom panel: Same as the middle panel but using the NICER results from Miller {\it et al.}~\cite{Miller19,Miller21}.}
	\label{fig:nsmrcompa}
\end{figure}

The mass and radius of neutron stars are obtained by solving the TOV equations for nonrotating stars. Figure~\ref{fig:nsmrcompa} shows the 95\% credibility regions for the mass $M$ and radius $R$ generated from the multivariate normal distribution for the EDF EOSs based on an ensemble of $\approx10^5$ EOSs. The top panel shows the prior distribution for the $k_{\rm F}$ expansion using different values of $d=1,3,5,7$, and $d=\infty$. The middle and lower panels show the posterior distribution including astrophysics information from GW170817 and the NICER analysis of Riley {\it et al.}~\cite{Riley19,Riley21} or the NICER analysis of Miller {\it et al.}~\cite{Miller19,Miller21}, respectively. Our results show that the posterior distributions obtained from the two different NICER analyses are very similar once the nuclear physics information is encoded in the EOS framework.

Regarding the different EDF choices, we find that the $d=3$ distribution is similar to the case of $d=5, 7,$ and $d=\infty$. However, large $d$, and in particular $d=\infty$ allows for the speed of sound to become acausal, $c_s^2 > 1$ (in units with the speed of light $c=1$), as the density increases, which is not the case in either neutron or symmetric matter for $d=3$ by construction. In addition, as $d=1$ makes the interaction energy density rapidly behave like $n^{5/3}$, the EOS becomes soft at rather low densities compared to the larger $d$ values~\cite{Huth21}. As a result the 95\% credibility regions for mass and radius only extend slightly above $2\msun$. Therefore, in the following, we will show results only for the EDF EOSs with $d=3$. Before doing so, we also list the radius ranges of typical $1.4\msun$ and $2\msun$ neutron stars to show the rather minor sensitivity to the choice of $d$ (see Table~\ref{tb:rad14}).

\begin{table}[t]
	\begin{tabular}{cccccc}
    \hline\hline 
    \multirow{2}{*}{EDF EOS} & $R_{-2\sigma}$ & $R_{-1\sigma}$ & $R_{\rm most}$ & $R_{+1\sigma}$ & $R_{+2\sigma}$ \\
    & ~~(km)~~ & ~~(km)~~ & ~~(km)~~ & ~~(km)~~ & ~~(km)~~ \\
    \hline\hline
    \multicolumn{6}{c}{Prior} \\
    \hline
    $d=1$ & 10.01 & 11.20 & 12.25 & 12.61 & 13.06 \\
    $d=3$ & 9.87 & 11.39 & 12.45 & 12.77 & 13.19 \\
    $d=5$ & 9.88 & 11.46 & 12.48 & 12.80 & 13.20 \\
    $d=7$ & 9.90 & 11.47 & 12.48 & 12.81 & 13.22 \\
    $d=\infty$ & 9.97 & 11.51 & 12.50 & 12.83 & 13.23 \\
    \hline\hline
    \multicolumn{6}{c}{Posterior Riley {\it et al.}} \\
    \hline
    $d=1$ & 12.00 & 12.40 & 12.78 & 13.02 & 13.28 \\
    $d=3$ & 11.57 & 12.13 & 12.60 & 12.87 & 13.17 \\
    $d=5$ & 11.48 & 12.08 & 12.58 & 12.86 & 13.16 \\
    $d=7$ & 11.39 & 12.06 & 12.58 & 12.85 & 13.16 \\
    $d=\infty$ & 11.33 & 12.02 & 12.58 & 12.85 & 13.16 \\
    \hline\hline
    \multicolumn{6}{c}{Posterior Miller {\it et al.}} \\
    \hline
    $d=1$ & 11.88 & 12.32 & 12.73 & 12.97 & 13.25 \\
    $d=3$ & 11.65 & 12.19 & 12.68 & 12.93 & 13.23 \\
    $d=5$ & 11.63 & 12.19 & 12.65 & 12.92 & 13.22 \\
    $d=7$ & 11.61 & 12.18 & 12.68 & 12.93 & 13.23 \\
    $d=\infty$ & 11.52 & 12.16 & 12.65 & 12.92 & 13.22 \\
    \hline\hline
	\end{tabular}
    \caption{Prior and posterior results for the radius of a 1.4$\msun$ neutron star at $\pm 2\sigma$, $\pm 1\sigma$, and the most likely radius for the EDF EOS ensembles with the $k_{\rm F}$ expansion and different $d$ values. Results are given for both NICER analyses (Riley {\it et al.}~\cite{Riley19,Riley21} and Miller {\it et al.}~\cite{Miller19,Miller21}).}
    \label{tb:rad14}.
\end{table}

\begin{table}[t]
	\begin{tabular}{lccccc}
    \hline\hline 
    \multirow{2}{*}{EDF EOS} & $R_{-2\sigma}$ & $R_{-1\sigma}$ & $R_{\rm most}$ & $R_{+1\sigma}$ & $R_{+2\sigma}$ \\
    & ~(km)~ & ~(km)~ & ~(km)~ & ~(km)~ & ~(km)~ \\
    \hline\hline 
    $d=3$ Prior & 10.46 & 11.15 & 12.03 & 12.40 & 12.84 \\
    \hline
    $d=3$ Post.~Riley & 10.86 & 11.56 & 12.15 & 12.43 & 12.75 \\
    \hline
    $d=3$ Post.~Miller & 10.93 & 11.63 & 12.25 & 12.51 & 12.82 \\
    \hline\hline 
	\end{tabular}
    \caption{Same as Table~\ref{tb:rad14} but for the radius of a 2.0$\msun$ neutron star and for the case of $d=3$ only.}
    \label{tb:rad20}.
\end{table}

\begin{figure}[t!]
   \centering
	\includegraphics[width=0.85\columnwidth]{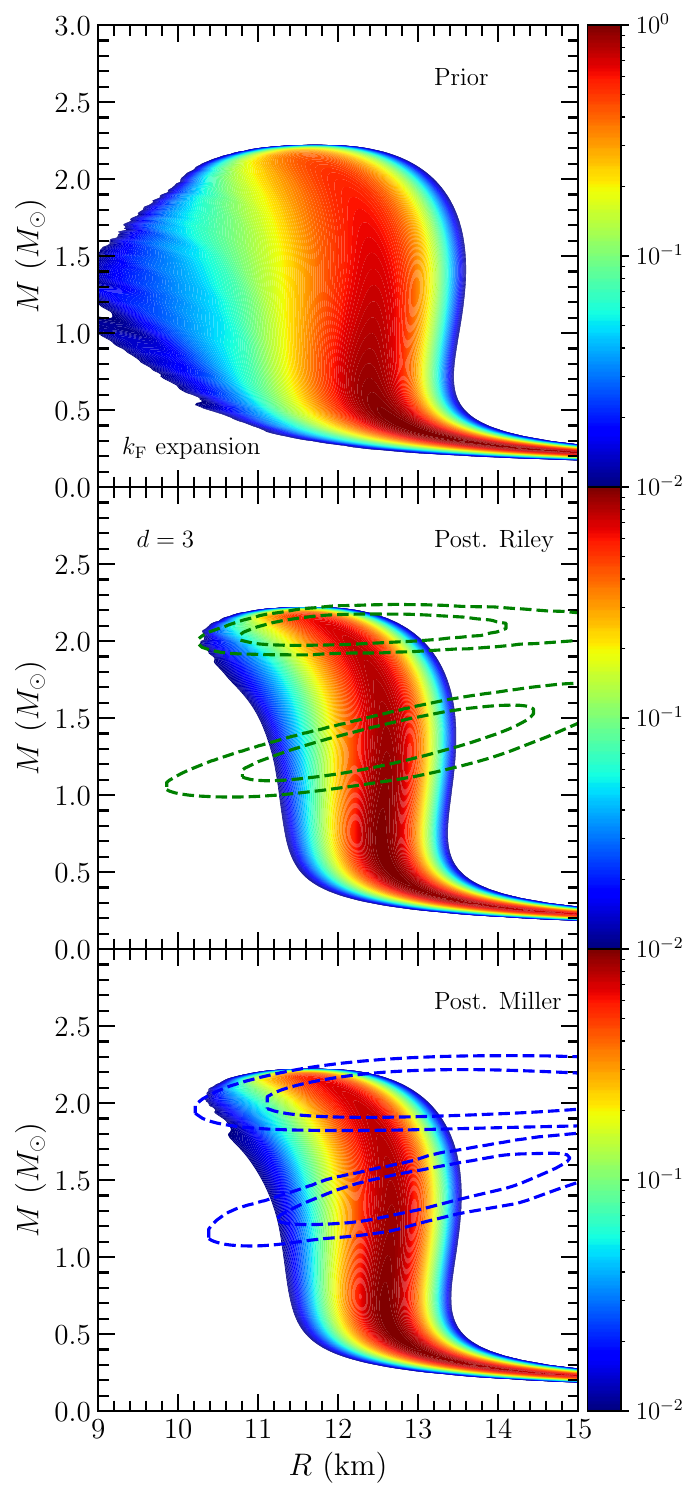}
	\caption{Same as Fig.~\ref{fig:nsmrcompa} but showing the color-coded prior and posterior distributions for the case of $d=3$.}
	\label{fig:nsmrcomp}
\end{figure}

In Table~\ref{tb:rad14} we give the prior and posterior ranges for the radius $R_{1.4}$ of a 1.4$\msun$ neutron star at 95\% ($\pm 2\sigma$) and 68\% ($\pm 1\sigma$) credibilities as well as the most likely radius for the EDF EOS ensembles with the $k_{\rm F}$ expansion and different $d$ values. For $d=3$, the 95\% credibility prior range is $R_{1.4} = (9.87\text{--}13.19)$\,km. Including the astrophysics information from GW170817 and the NICER analysis of Riley {\it et al.}~\cite{Riley19,Riley21} gives for 95\% credibility posterior range $R_{1.4}=(11.57\text{--}13.17)$\,km, while with the Miller {\it et al.}~\cite{Miller19,Miller21} analysis $R_{1.4}=(11.65\text{--}13.23)$\,km, or the combined range $R_{1.4}=(11.6\text{--}13.2)$\,km. Both NICER analyses thus give very similar posterior ranges with the result based on Miller {\it et al.} shifted to slightly larger radii. Overall, the radius range decreases by over 50\% from 3.3\,km for the prior to 1.6\,km for the combined posterior, mainly by disfavoring the smaller radii in the prior range. Moreover, in the prior distribution for $d=3$, 72\% of EOSs have a maximum mass of neutron stars greater than $2.0\msun$, 
while for the posterior distribution, 97\% (98\%) of EOSs have a maximum mass above $2.0\msun$ using
the NICER analysis of Riley {\it et al.}~\cite{Riley21} (Miller {\it et al.}~\cite{Miller21}).

In Fig.~\ref{fig:nsmrcomp},we show the color-coded prior and posterior distributions for the case of $d=3$. In both posterior distributions, the most probable radii for neutron stars between $1.0\msun$ and $1.8\msun$ vary only within $0.3\,\rm{km}$. Moreover, the mass and radius distribution for $M>2.0\msun$ is very similar for the prior and the two posteriors, 
because the astrophysics information mainly removes EOSs that give low maximum mass and small radii.

Table~\ref{tb:rad20} gives the prior and posterior ranges for the radius $R_{2.0}$ of a 2.0$\msun$ neutron star for the EDF EOSs with $d=3$. The prior distribution shows a wider radius range because it does not include information of a massive neutron star. Again the two posterior ranges for $R_{2.0}$ are very similar and merely shifted by less than 100\,m. In the case of $d=3$, the maximum mass of neutron stars among the $\sim 10^5$ EOS ensemble reaches up to $2.23\msun$, while it can go up to $2.32\msun$ for $d=\infty$.

\subsection{Symmetry energy and $L$ parameter}

\begin{figure}[t]
	\centering
	\includegraphics[width=0.9\columnwidth]{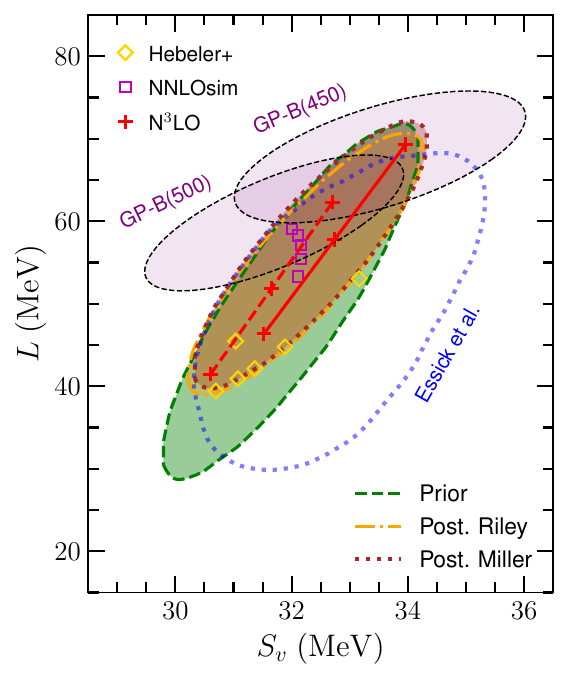}
	\caption{Plot of the $L$ parameter versus symmertry energy $S_v$ based on the MBPT calculations for different chiral NN+3N Hamiltonians from Ref.~\cite{Dris19MCshort} [points, see upper left legend; the dashed (solid) line connects the 500 (450)\,MeV cutoff N$^3$LO results] and the resulting prior and posterior distributions based on the EDF EOS ensemble of this work for the $k_{\rm F}$ expansion and $d=3$ (see lower right legend). Note that the posteriors for both NICER analyses are very similar. For comparison, we also plot the GP-B results at N$^3$LO from Ref.~\cite{Dris20PRL} and the recent results from Essick {\it et al.}~\cite{Essick21PRC}. All contours are 95\%, except for 90\% from Essick {\it et al.} Note that all results are for a fixed reference saturation density $n_0=0.16$\,fm$^{-3}$, while the GP-B results are for the correlated range of 95\% of the calculated saturation density.}
	\label{fig:svl}
\end{figure}

We can also extract the symmetry energy $S_v$ and the slope parameter $L$ from our calculations. To this end, we calculate the symmetry energy parameters from our EDF EOS ensemble of $(a_i,b_i)$ generated by the covariance matrix $C_{jk}$ [see Eq.~\eqref{eq:cov}] from
\begin{equation}
    S_v = \frac{1}{8}\frac{\pt^2}{\pt x^2}\biggl(\frac{\varepsilon}{n}\biggr) \,,\quad
    L = \frac{3}{8} \, n \, \frac{\pt^3}{\pt x^2\pt n}\biggl(\frac{\varepsilon}{n}\biggr) \,,
\end{equation}
where both symmetry parameters are evaluated at $n=n_0$ and $x=\frac{1}{2}$. The resulting distributions for $S_v$ and $L$ follow Gaussian distributions, with the mean and covariance matrix given in the following equation. This is shown in Fig.~\ref{fig:svl} for the individual MBPT calculations for the different chiral NN+3N Hamiltonians from Ref.~\cite{Dris19MCshort} as points, where the dashed (solid) line connects the 500 (450)\,MeV cutoff N$^3$LO results. As discussed, our EDF EOS ensembles are built from all the different chiral NN+3N results. The resulting $95\%$ prior and posterior distributions are shown for the EDF EOS ensemble with the $k_{\rm F}$ expansion and $d=3$. 
Note that our $95\%$ distribution contours are obtained by integrating the 2D (two-dimensional) domain with 2D probability for $S_v$ and $L$.
We find that the prior range for $S_v$ and $L$ is narrowed to larger values with the astrophysics constraints included. For both NICER analyses the posteriors are again very similar.

For the prior distribution these are given by (mean values in MeV and convariance matrix in MeV$^2$):
\begin{equation}
\langle S_v, L \rangle = (31.96, 51.70) \,, \quad
\Sigma_{S_v,L} = 
\begin{pmatrix}
0.79 & 6.73 \\
6.73 & 75.11
\end{pmatrix}
\,,
\end{equation}
while the posterior distributions for the astrophysical inferences are given for the Riley {\it et al.}~\cite{Riley21} and Miller {\it et al.}~\cite{Miller21} analyses, respectively:
\begin{equation}
\langle S_v, L \rangle = (32.23, 56.33) \,, \quad
\Sigma_{S_v,L}^{\rm Riley}= 
\begin{pmatrix}
0.66 & 4.56 \\
4.56 & 40.02
\end{pmatrix}
\,,
\end{equation}
and
\begin{equation}
\langle S_v, L \rangle = (32.31, 57.31) \,, \quad
\Sigma_{S_v,L}^{\rm Miller} = 
\begin{pmatrix}
0.64 & 4.43 \\
4.43 & 40.43
\end{pmatrix}
\,.
\end{equation}
We observe that the astrophysics constraints move the posterior distributions to larger $S_v$ and $L$ values within the prior range. Moreover, all MBPT calculations for the different chiral $NN+3N$ Hamiltonians are still largely within the posterior range, but some of them only borderline. This points to that astrophysics prefers EOSs on the stiffer part of the neutron matter EOS band based on chiral EFT. This is consistent with the EOS findings in Ref.~\cite{Raaijmakers21}.

In Fig.~\ref{fig:svl} we also show the GP-B (Gussian process, BUQEYE Collaboration) results at N$^3$LO from Ref.~\cite{Dris20PRL}. Since the GP-B contours are based on the same N$^3$LO 500 (450)\,MeV results~\cite{Dris19MCshort} included in our analysis, we can trace the difference between the GP-B contours and the N$^3$LO points to the evaluation of $S_v$ and $L$ for the correlated range of 95\% of the calculated saturation density, while our distributions are at a fixed reference saturation density $n_0=0.16$\,fm$^{-3}$. Since the $L$ parameter scales linearly with the density, this mainly affects the $L$ value, while the range of symmetry energies is broadened due to the additional uncertainty in the calculated saturation density.

Finally, we compare our $95\%$ posterior distributions in Fig.~\ref{fig:svl} with the recent results from Essick {\it et al.}~\cite{Essick21PRC}, which are however $90\%$ contours.
These are based on a different set of chiral $NN+3N$ calculations and astrophysics constraints through a more general Gaussian process extension to high densities. 
Nevertheless both contours (at the same reference saturation density $n_0$) are remarkably consistent.

\subsection{Proton fraction}
\label{protonfraction}

The ground state of neutron star matter is obtained by solving the condition for beta equilibrium, 
\begin{equation}
\mu_n = \mu_p + \mu_e \,,
\label{eq:beta}
\end{equation}
where the neutron, proton, and electron chemical potentials $\mu_n$, $\mu_p$, and $\mu_e$ are given by
\begin{equation}
\mu_n = \frac{\pt \varepsilon}{\pt n_n},\, \quad
\mu_p = \frac{\pt \varepsilon}{\pt n_p},\, \quad
\mu_e = \frac{\pt \varepsilon}{\pt n_e}\,,
\end{equation}
with total energy density $\varepsilon$. Since the core is composed of uniform nuclear matter, Eq.~\eqref{eq:beta} is straightforward for a given EDF. For the crust EOS, where matter exists in inhomogeneous form, we employ the liquid drop model (LDM)~\cite{Lim17} using the same EDF to construct the EOSs of the inner and outer crust.

In the inner crust, the total energy density including the electron contribution is given by~\cite{Lim17}
\begin{multline}
\varepsilon = u n_i f_i + \frac{\sigma(x_i)u d}{r_N}
+ 2\pi (e x_i n_i r_N)^2  u f_d(u) \\[1mm]
+ (1-u)n_{no}f_{no} + \varepsilon_e \,,
\end{multline}
where $u$ is the volume fraction of the nucleus to the Wigner-Seitz cell, 
$n_i$  is the baryon number density of the heavy nucleus,
$n_{no}$ is the density of unbound neutrons, 
$x_i$ is the proton fraction in the heavy nucleus, and
$f_i =f(n_i,x_i)$ and $f_{no}=f(n_{no},x_{no}=0)$ are the energies per baryon for the heavy nucleus and unbound neutrons, respectively.
$\sigma(x_i)$  is the surface tension at zero temperature as a function of the proton fraction in heavy nuclei, $r_N$ the radius of the heavy nucleus, $e$ the electric charge, $d$ the dimension of the nuclear pasta phase, $f_d(u)$ the Coulomb shape function corresponding to the nuclear pasta phase, and $\varepsilon_e$ is the electron energy density.
We use the surface tension from~\cite{RPL1983}
\begin{equation}
\sigma(x_i) = \sigma_0 \, \frac{2^{\alpha+1}+ q}{x^{-\alpha} + q + (1-x)^{-\alpha}} \,,
\end{equation}
where $\sigma_0$, $\alpha$, and $q$ are parameters fit to the calculation of the surface tension. In this work, we use $\sigma_0= 1.14\,\mathrm{MeV\,fm}^{-2}$, $\alpha=3.4$, and $q=30$, but note that the crust properties depend only weakly on the surface tension parameters, and also the impact of the crust on the investigated neutron star properties is minor.

Based on the viral theorem, the Coulomb energy is approximately twice the nuclear surface energy. Thus, we can combine the surface and Coulomb energy to a single form of energy contribution, which leads to a simpler equation for the energy density~\cite{LSEOS},
\begin{multline}
\varepsilon =  u n_i f_i + \left(\frac{243\pi}{5}e^2 x_i^2 n_i^2 \sigma^2(x_i) \right)^{1/3} \mathcal{D}(u) \\[1mm]
+ (1-u)n_{no}f_{no} + \varepsilon_e \,,
\label{eq:crust}
\end{multline}
where $\mathcal{D}(u)$ is a continuous dimension function introduced in Ref.~\cite{LSEOS}. For total baryon density and proton fraction $Y_p$, and thus electron density $n_e = Y_p n$, the conditions $u$, $n_i$, $x_i$, and $n_{no}$ are found by minimizing the total energy density, Eq.~\eqref{eq:crust}, using the Lagrange multiplier method for the constraints of baryon density and charge neutrality,
\begin{equation}
n = u n_i + (1-u) n_{no} \quad \text{and} \quad n_e = (1-u) n_i x_i \,.
\end{equation}
For an outer crust EOS, which is defined as the region without unbound neutrons, the outside neutron density $n_{no}$ is neglected. Using the LDM construction, the transitions from the outer to inner crust and to the outer core are thus smooth, since the same EDF is employed to construct the entire neutron star EOS.

\begin{figure}[t]
	\centering
	\includegraphics[width=\columnwidth]{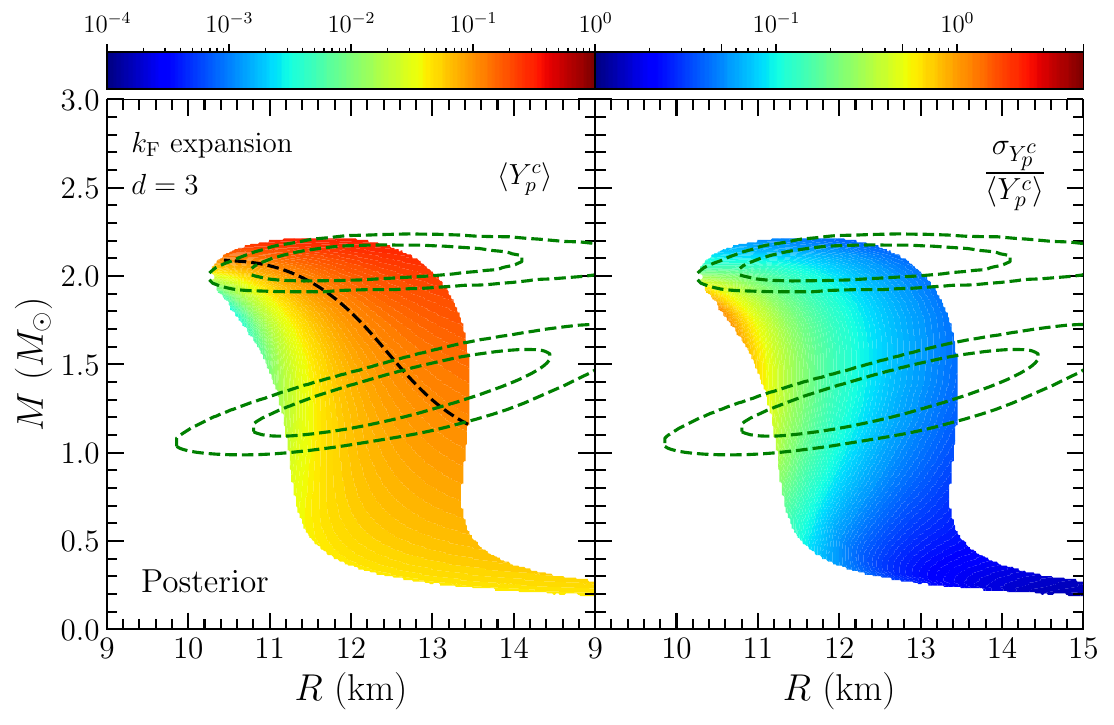}
	\caption{Color-coded posterior distribution of the proton fraction at the central density based on the EDF EOS ensemble for the $k_{\rm F}$ expansion and $d=3$ (using the NICER Riley {\it et al.}~analysis). The left and right panels show the average proton fraction at the central density $\langle Y_p^c \rangle$ and the variance over the average $\sigma_{Y_p^c} / \langle Y_p^c \rangle$, respectively. The black dashed line denotes the direct Urca threshold of $Y_p = 1/9$.}
	\label{fig:nsmryp}
\end{figure}

Figure~\ref{fig:nsmryp} shows the average proton fraction at the central density $\langle Y_p^c \rangle$ based on the EDF EOS ensemble for the $k_{\rm F}$ expansion and $d=3$, as well as the variance over the average $\sigma_{Y_p^c} / \langle Y_p^c \rangle$. The average proton fraction is dominated by the core, but include the details of the crust calculation discussed above. We note that in Fig.~\ref{fig:nsmryp} (and in Figs.~\ref{fig:nsmrrhoc} and~\ref{fig:nsmrspeed}),
the mass and radius domain is restricted to the region where the relative probability to the maximum probability ratio $P(M,R)/P_{\rm max} \geq 10^{-2}$ (as in Fig.~\ref{fig:nsmrcomp}). As expected, the proton fraction increases as the mass increases, and for a given mass, it increases with radius as the EOS becomes stiffer. Our EOS ensemble assumes for the proton fraction that matter is nucleonic, which may not be valid for massive stars. However, for typical $1.4 \msun$ neutron stars, this may not be such a large extrapolation.

In addition, we plot in Fig.~\ref{fig:nsmryp} the threshold  $Y_p = 1/9$ for the direct Urca process, which leads to fast cooling neutron stars~\cite{Lattimer91}. We find that typical neutron stars around $1.4 \msun$ do not exceed this threshold for radii around 12\,km, but only in our largest radius configurations. However, based on our results, we expect that massive neutron stars with $M>2.1\msun$  would cool via the direct Urca process.

\begin{figure}[t]
	\centering
    \includegraphics[width=\columnwidth]{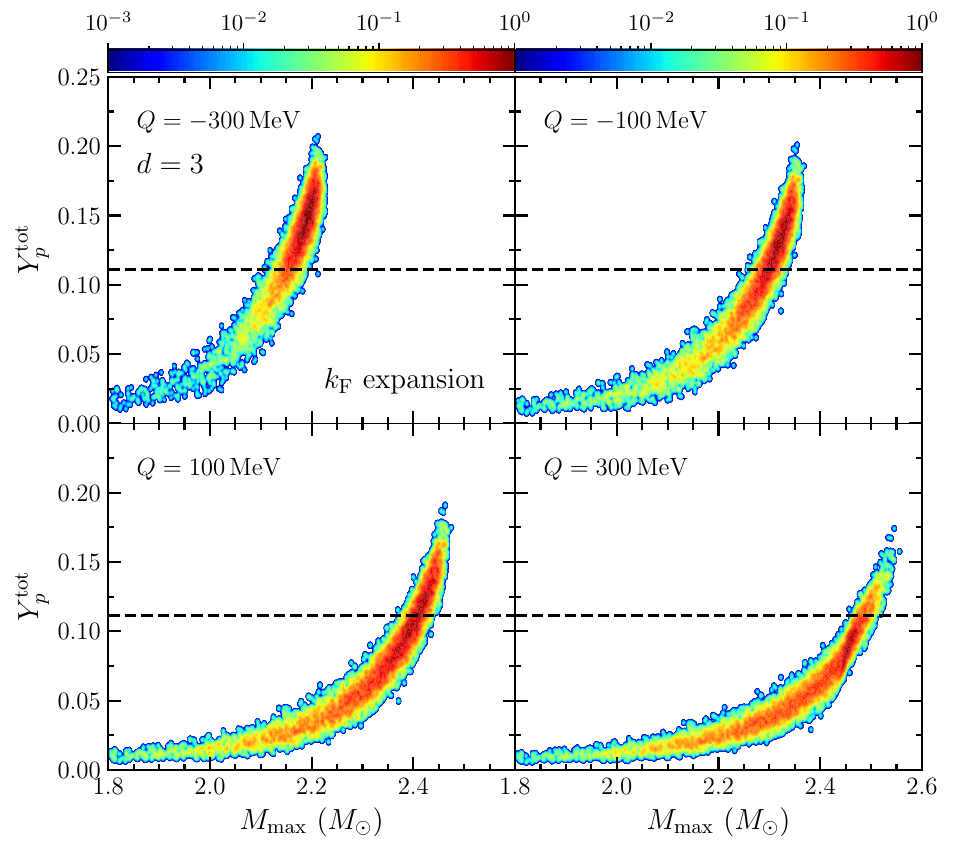}
	\caption{Color-coded posterior distribution of the total proton fraction $Y_p^{\rm tot}$ of the maximum mass star versus the maximum mass for four different $Q$ values (different panels) based on the EDF EOS ensemble for the $k_{\rm F}$ expansion and $d=3$ (using the NICER Riley {\it et al.}~analysis). The black dashed line denotes the direct Urca threshold of $Y_p = 1/9$.}
	\label{fig:massmaxyptot}
\end{figure}

\begin{figure}[t]
	\centering
	\includegraphics[width=\columnwidth]{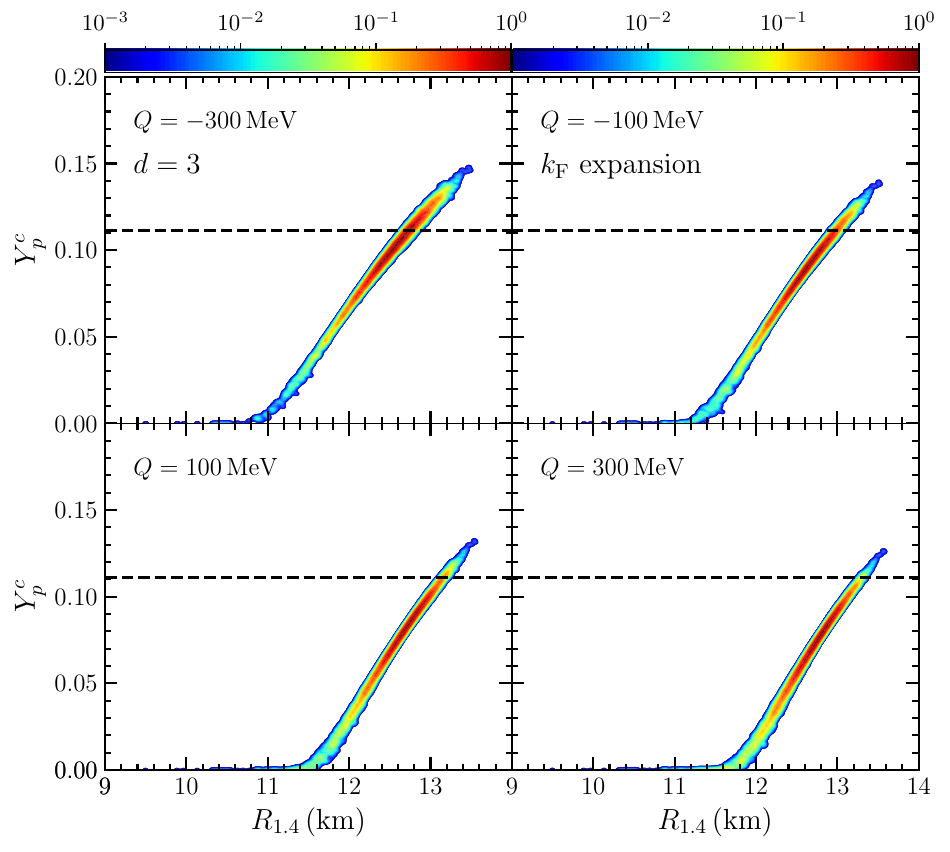}
	\caption{Same as Fig.~\ref{fig:massmaxyptot} but for the proton fraction at the central density $Y_p^c$ versus the radius of a $1.4 \msun$ neutron star.}
	\label{fig:mass14ypc}
\end{figure}

Figure~\ref{fig:massmaxyptot} shows the total proton fraction $Y_p^{\rm tot}$ of the maximum mass star versus the maximum mass. The total proton fraction increases along a band as the maximum mass increases, due to the stiffer EOS. Figure~\ref{fig:massmaxyptot} shows results for four different $Q$ values of symmetric nuclear matter, keeping in mind that negative $Q$ values are favored by nuclear masses, ab initio calculations, and astrophysics~\cite{Dutra12PRC,Somasundaram21,Essick21PRC}.
With increasing $Q$, the total proton fraction for a given mass decreases and also the maximum mass increases, as larger $Q$ stiffens the EOS.
Naturally, the sensitivity to $Q$ is much less pronounced for typical neutron stars.
Figure~\ref{fig:mass14ypc} shows the proton fraction at the central density $Y_p^c$ versus the radius of a $1.4 \msun$ star, which exhibits a tight correlation and is only very weakly dependent on $Q$. Larger radii thus have a larger proton fraction. Again we see that radii around 12\,km, as expected based on most recent EOS astrophysical inferences~\cite{Lim19,Capano20,Al-Mamun21,Miller21,Raaijmakers21,Essick21,Huth22,Annala22,Altiparmak22,Gorda22},
do not cool via the direct Urca threshold. However for larger radii $R_{1.4} > 12.6$\,km (for $Q=-300$\,MeV) even typical neutron stars would be fast coolers.

\subsection{Central density and speed of sound}

\begin{figure}[t]
	\centering
	\includegraphics[width=\columnwidth]{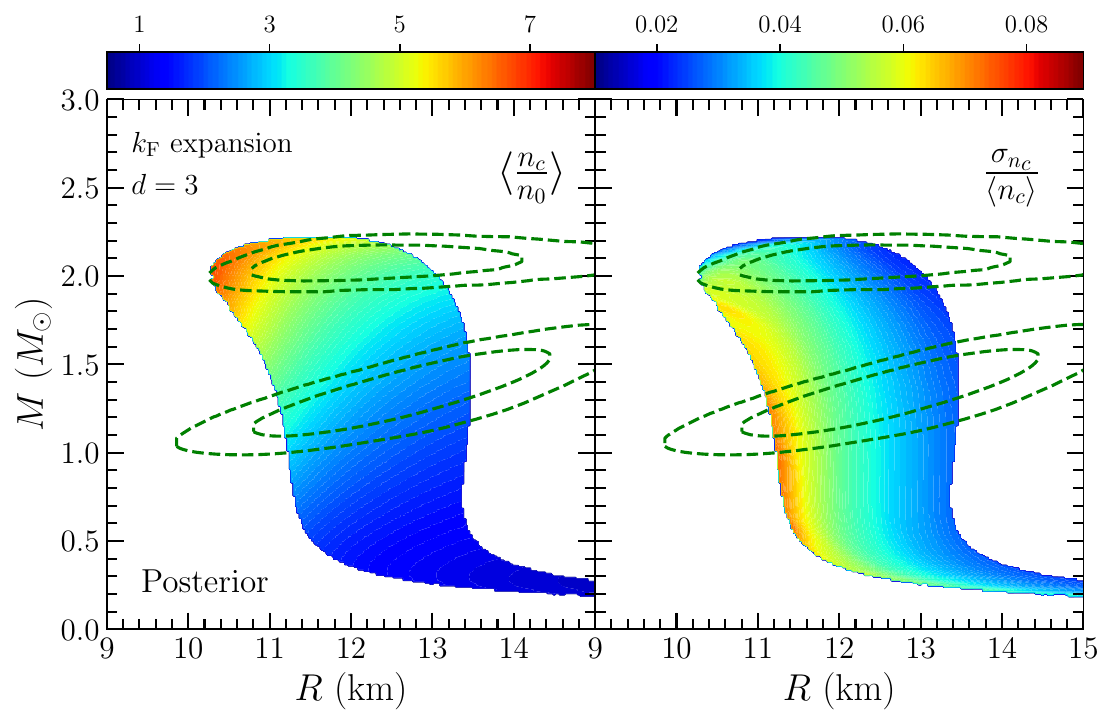}
	\caption{Same as Fig.~\ref{fig:nsmryp} but for the average central density in units of saturation density $\langle n_c/n_0 \rangle$ (left panel) and its variance over the average $\sigma_{n_c} / \langle n_c \rangle$ (righ panel).}
	\label{fig:nsmrrhoc}
\end{figure}

Next, we study the posterior distribution for the central density and the speed of sound in neutron stars. Figure~\ref{fig:nsmrrhoc} shows the average central density in units of saturation density, $\langle n_c/n_0 \rangle$, and its variance over the average, $\sigma_{n_c} / \langle n_c \rangle$. The average central density increases with increasing mass, while it decreases as the radius increases for a given mass of neutron star. This results from stiffer EOSs leading to larger radii. In our EDF EOSs, the maximal central density reaches up to $\approx 7 n_0$, which is reached for softer EOSs in the most massive neutron stars with smaller radii.

\begin{figure}[t]
	\centering
	\includegraphics[width=\columnwidth]{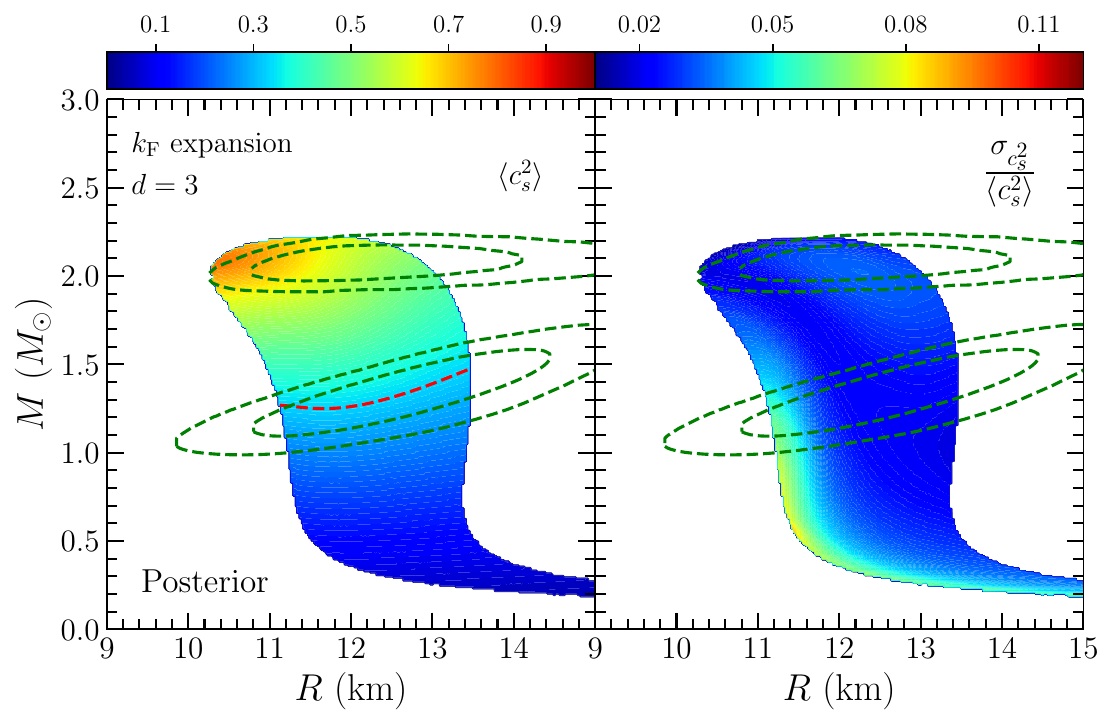}
	\caption{Same as Fig.~\ref{fig:nsmrrhoc} but for the square of the speed of sound $c_s^2$ at the central densities in neutron stars. The red dashed line represents the conformal limit $c_s^2 =1/3$.}
	\label{fig:nsmrspeed}
\end{figure}

Figure~\ref{fig:nsmrspeed} shows the speed of sound squared $c_s^2 = \partial P/\partial \varepsilon$ at the central densities in neutron stars. In our EDFs, the speed of sound increases but remains causal and decreases at high density~\cite{Huth21}. As we see from Fig.~\ref{fig:nsmrspeed}, the speed of sound is increasing as the mass increases, so in neutron stars most matter is on the part of the EOS that has an increasing $c_s^2$ in our ensemble of EOSs. In Fig.~\ref{fig:nsmrspeed}, the red dashed line represents $c_s^2=1/3$, which shows that even typical $1.4 \msun$ stars exceed the conformal limit, except when they have radii larger than 13\,km (see also the middle panel of Fig.~\ref{fig:mass14}). Moreover, information on the radii of massive stars with $M \gtrsim 2.0\msun$ would inform us about $c_s^2$ at the central density (see also Fig.~\ref{fig:mass20radcs2}). This could be realized with an improved NICER radius measurement~\cite{Riley21,Miller21} of the $2.08 \pm 0.07 \msun$ pulsar PSR J0740+6620~\cite{Fonseca21}.

\subsection{Correlations}

\begin{figure}[t!]
	\centering
	\begin{minipage}{\columnwidth}
		\centering
		\includegraphics[width=0.9\columnwidth]{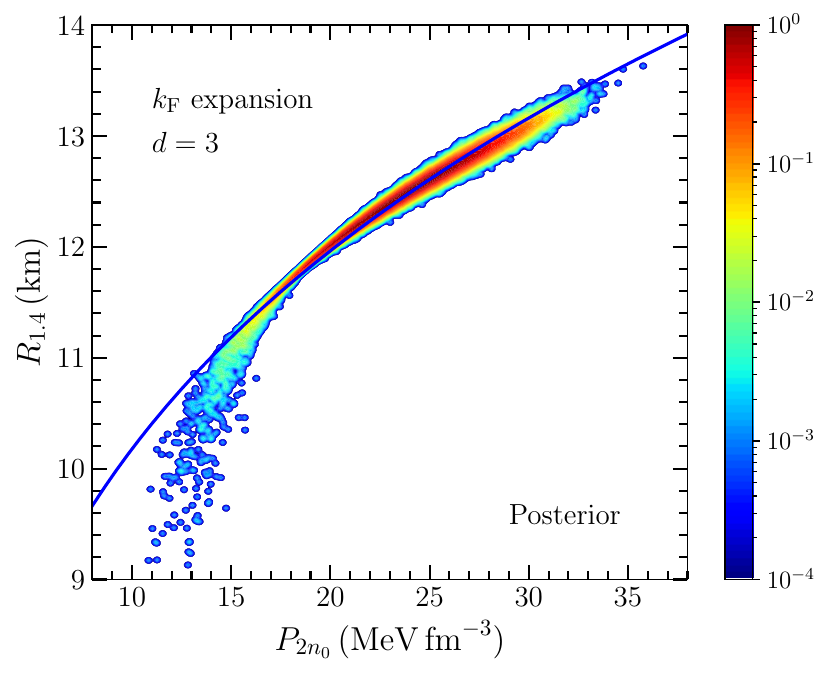}
	\end{minipage}\hfill
	\begin{minipage}{\columnwidth}
		\centering
		\includegraphics[width=0.9\columnwidth]{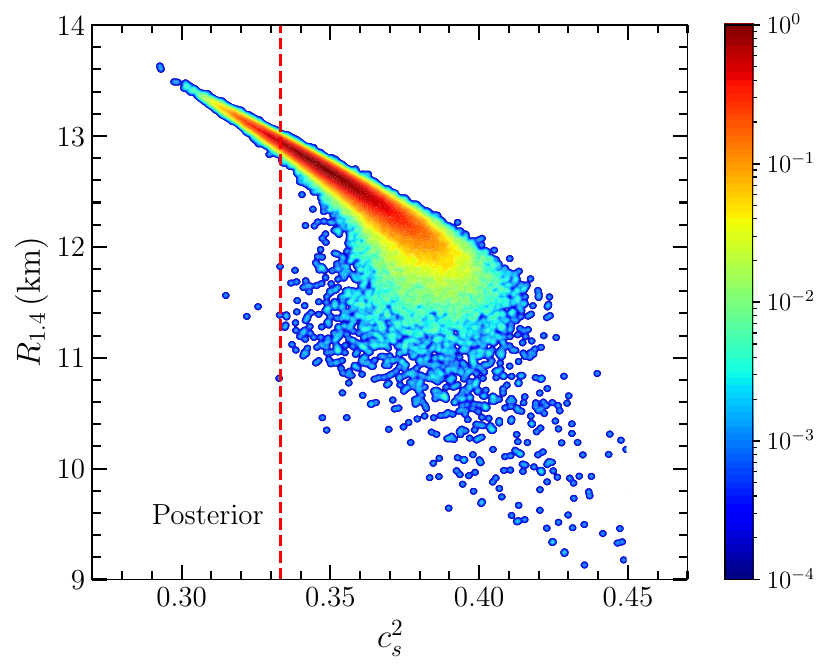}
	\end{minipage}
    \begin{minipage}{\columnwidth}
		\centering
		\includegraphics[width=0.9\columnwidth]{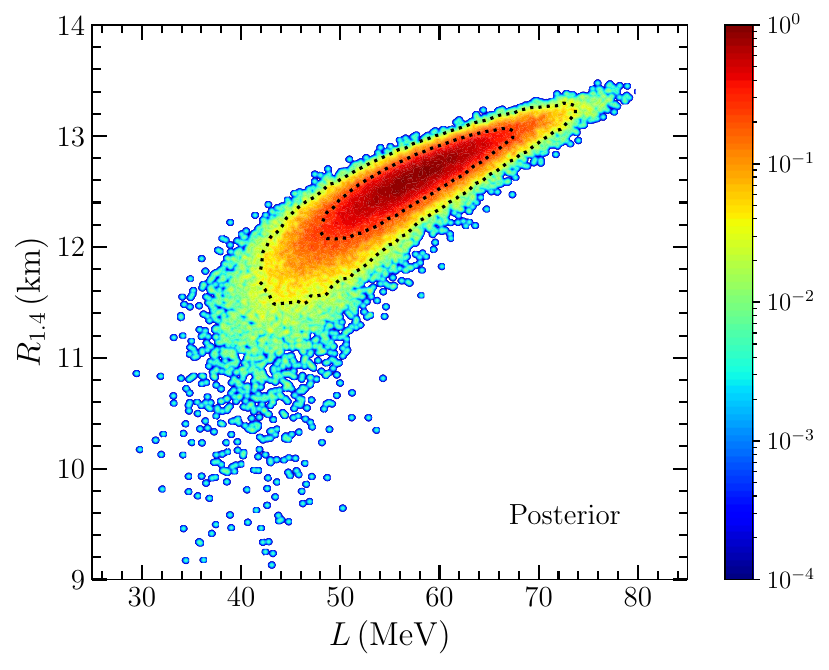}
	\end{minipage}
	\caption{Radius of a 1.4$\msun$ neutron star versus the pressure at twice saturation density $P_{2n_0}$ (top), the speed of sound squared at the central density $c_s^2$ (middle), and the $L$ parameter (bottom panel) based on the EDF EOS ensemble for the $k_{\rm F}$ expansion and $d=3$ (using the NICER Riley {\it et al.}~analysis). Recall that the $L$ parameter is proportional to the pressure of pure neutron matter at saturation density. In the middle panel the red dashed line represents the conformal limit $c_s^2 =1/3$.}
	\label{fig:mass14}
\end{figure}

Finally, we study the correlation of neutron star radii with the pressure and the speed of sound. In Ref.~\cite{Lattimer06} it was suggested that the radius of a $1.4\msun$ neutron star would follow the emprical relation $R_{1.4} \sim p_{2n_0}^{1/4}$, where $p_{2n_0}$ is the pressure at twice saturation density. In the top panel of Fig.~\ref{fig:mass14} we show that this correlation is indeed fulfilled in our EDF EOS ensemble within a band. For the radius in km and the pressure in MeV\,fm$^{-3}$, we find $R_{1.4} = 1.279 + 5.063 \, P_{2n_0}^{1/4}$ for the mean line of the correlation shown in Fig.~\ref{fig:mass14}, with a correlation coefficient $r_{xy} = 0.985$. While the details of this correlation depend on the EOS model, this indicates that astrophysical observations of neutron star radii provide constraints for the pressure at twice saturation density.

The middle panel of Fig.~\ref{fig:mass14} shows the distribution of $R_{1.4}$ versus the speed of sound at the central density of neutron stars. Most of the distribution follows a linear trend, but the correlation coefficient $r_{xy}=-0.907$ is weaker in this case. We also observe that $c_s^2$ at the central density exceeds the conformal limit $c_s^2 =1/3$ in our EDF EOS ensemble for $R_{1.4}$ smaller than 12.8\,km The correlation is even weaker at lower densities when comparing $R_{1.4}$ with the $L$ parameter in the bottom panel of Fig.~\ref{fig:mass14}, which is proportional to the pressure of pure neutron matter at saturation density. This is as expected because the central density of a $1.4\msun$ neutron star is $\sim 3 n_0$. Nevertheless, there is a general trend that $R_{1.4}$ increases as $L$ increases. 

\begin{figure}[t]
    \centering
    \includegraphics[width=0.9\columnwidth]{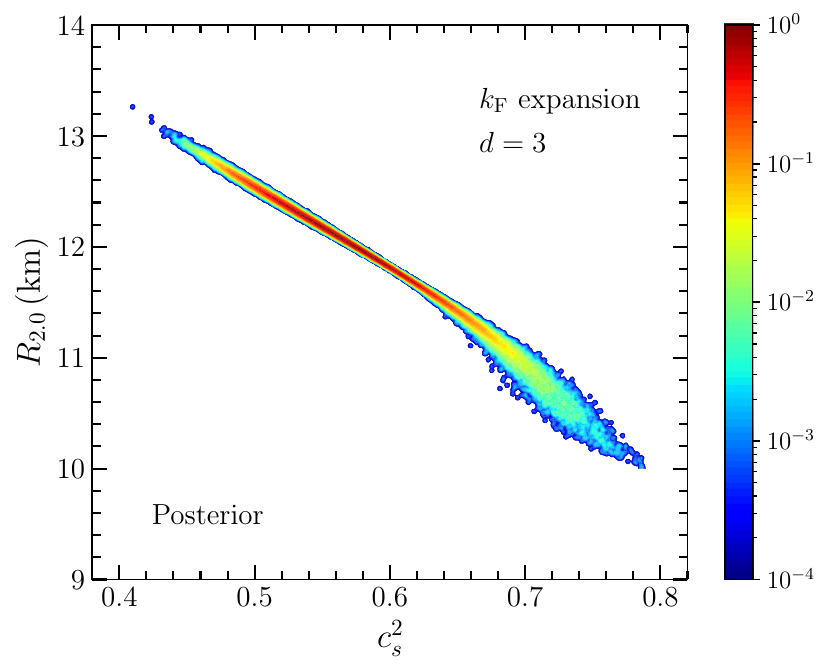}
    \caption{Same as the middle panel of Fig.~\ref{fig:mass14} but for the radius of a 2.0$\msun$ neutron star.}
    \label{fig:mass20radcs2}
\end{figure}

\begin{table}[b]
    \centering
\caption{Fitting functions for neutron star radii for different nuclear matter properties and the corresponding correlation coefficients (as discussed in the text).}
    \begin{tabular}{l|r}
    \hline
    \hline
    Fitting function & $r_{xy}$ \\
    \hline
    \hline
$R_{1.4} = 1.279 + 5.063 \, P_{2n_0}^{1/4}$ \: & \: $0.985$ \\
$R_{1.4} = 19.550 -19.757 \, c_s^2(n_c)$ \: & \: $-0.907$ \\
$R_{2.0} = 16.493 - 7.846 \, c_s^2(n_c)$ \: & \: $-0.996$ \\
    \hline
    \hline
    \end{tabular}
    \label{tab:corrfit}
\end{table}

Figure~\ref{fig:mass20radcs2} shows the correlation of the radius of a 2.0$\msun$ neutron star with the speed of sound at the central density. The strong correlation indicates that the radius measurement of massive neutron stars provides constraints for the speed of sound in dense nuclear matter. For the radius in km, we find $R_{2.0} = 16.493 - 7.846 \, c_s^2(n_c)$, with a correlation coefficient $r_{xy}=-0.996$. Moreover, we find within our EDF EOS ensemble that the speed of sound at the central density of 2.0$\msun$ stars is always greater than the conformal limit. Table~\ref{tab:corrfit} summarizes the fitting functions for neutron star radii for different nuclear matter properties and the corresponding correlation coefficients discussed above.

\begin{figure}[t]
	\centering
	\begin{minipage}{\columnwidth}
		\centering
		\includegraphics[width=0.9\columnwidth]{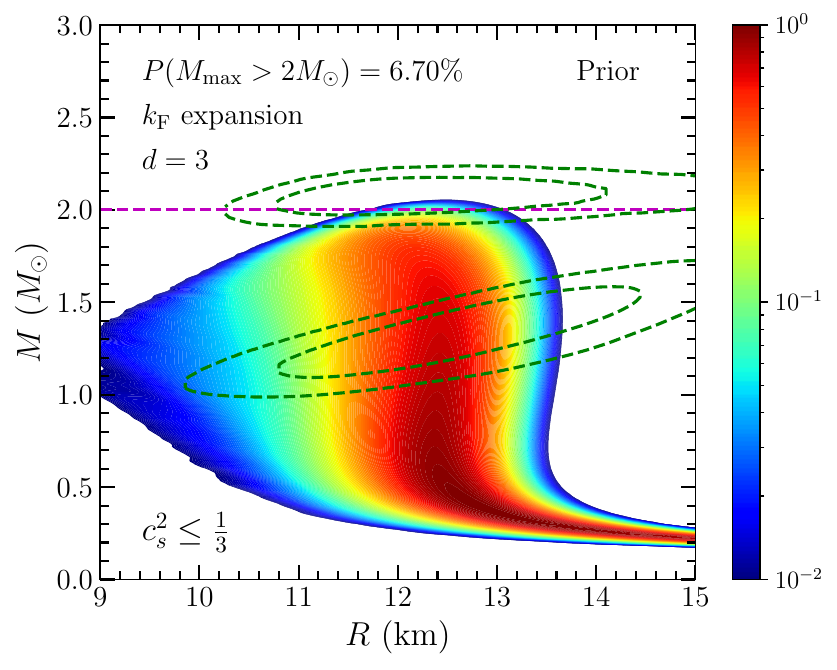}
	\end{minipage}\hfill
	\begin{minipage}{\columnwidth}
		\centering
		\includegraphics[width=0.9\columnwidth]{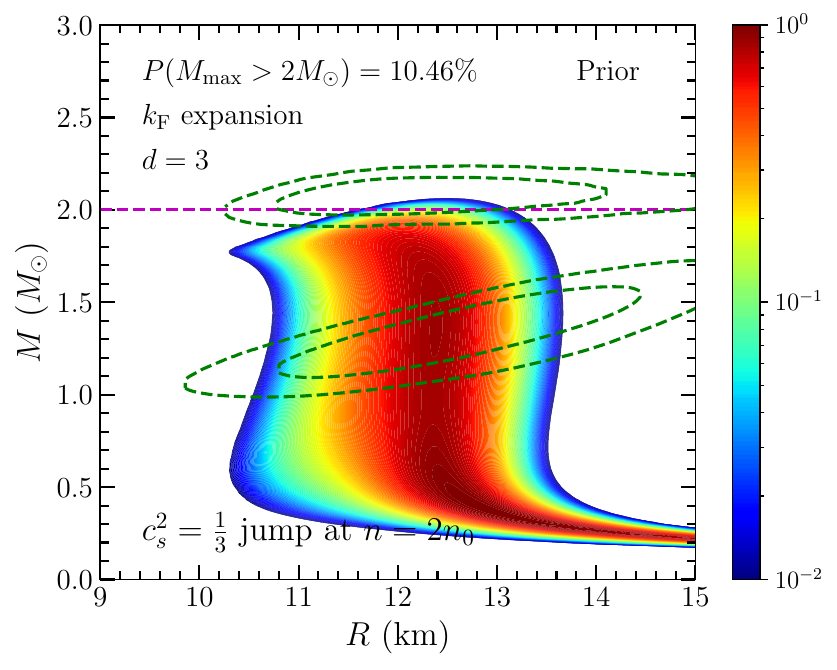}
	\end{minipage}
	\caption{Neutron star mass and radius prior based on the EDF EOS ensemble for the $k_{\rm F}$ expansion and $d=3$ when we impose the conformal limit of the speed of sound squared $c_s^2 \leq 1/3$ (top panel) or a jump in the speed of sound squared to constant $c_s^2=1/3$ at $n=2n_0$ (bottom panel).}
	\label{fig:cs213}
\end{figure}

Figure~\ref{fig:cs213} shows the mass and radius prior when we impose the conformal limit for the speed of sound. The top panel shows the case when the speed of sound continues to increase up to $1/3$ and maintains the conformal limit for all higher densities. The bottom panel is for the case where the speed of sound jumps to $1/3$ at $n=2n_0$ and remains at the conformal limit for all higher densities. In both scenarios, the speed of sound is not larger than the conformal limit at any density. From Fig.~\ref{fig:cs213}, the prior probability of supporting $2.0\msun$ stars is around $10\%$ or less, which is similar to the findings of Ref.~\cite{Bedaque15}. Thus, the conformal limit can be consistent with $2.0\msun$ stars, but most of the support of our EDF EOS ensemble exceeds the conformal limit for massive neutron stars. However, when we take the maximum mass limit as the central value of PSR J0740+6620, $2.08\msun$, the speed of sounds needs to exceed $1/3$ in our ensemble, as the maximum mass does not reach up to $2.08\msun$ in our modeling in both cases in Fig.~\ref{fig:cs213}.

\section{Summary and conclusion}
\label{sec:summary}

We have explored EOS ensembles using new EDFs from Ref.~\cite{Huth21} that allow for large variations at high densities. The EDF EOS ensembles were constrained by empirical properties of symmetric nuclear matter and by MBPT calculations of neutron matter based on different chiral $NN+3N$ Hamiltonians. Starting from this prior, constraints at high densities were included from observations of GW170817 and from NICER, where the heavy neutron star mass constraint is incorporated through PSR J0740+6620. All our results show that both the Riley {\it et al.}~\cite{Riley21} and Miller {\it et al.}~\cite{Miller21} NICER analyses lead to very similar posterior constraints for the symmetry energy and neutron star properties when folded into our EOS framework.

Based on our EDF EOS ensembles, we have studied the symmetry energy and the $L$ parameter, as well as the proton fraction, the speed of sound, and the central density in neutron stars. Our 95\% posterior credibility ranges for the symmetry energy $S_v$, the $L$ parameter, and the radius of a 1.4$\msun$ neutron star $R_{1.4}$ are $S_v=(30.6\text{--}33.9)$\,MeV, $L=(43.7\text{--}70.0)$\,MeV, and $R_{1.4}=(11.6\text{--}13.2)$\,km. Moreover, we have shown that larger and/or heavier neutron stars have a larger proton fraction and are thus more likely to cool rapidly via the direct Urca process. 

As can be seen from our results for $S_v$ and $L$, present astrophysics constraints prefer larger pressures within the prior ranges. To this end, we have also explored correlations of neutron star radii with the pressure and the speed of sound. The radius of $1.4\msun$ stars was found to correlate well with the pressure at twice saturation density, and $R_{2.0}$ was shown to correlate tightly with the speed of sound at the central density. Therefore, precise measurements of $R_{1.4}$ provide key information for density regimes at the limits of chiral EFT calculations, and radii of massive neutron stars will help to constrain the behavior of the speed of sound in dense matter. Finally, by constructing EOS ensembles with imposed conformal limit on the speed of sound, we found that a maximum mass of neutron stars $M_{\rm max}>2.1\msun$ indicates that the speed of sound needs to exceed the conformal limit.

\acknowledgments

We thank Sabrina Huth for fruitful discussion. This work was supported by the Max Planck Society, the European Research Council (ERC) under the European Union's Horizon 2020 research and innovation programme (Grant Agreement No.~101020842) and by the National Research Foundation of Korea (NRF) Grant funded by the Korea government (MSIT) (No. 2021R1A2C2094378).


%

\end{document}